# Mini Review

**Running head**
Statistical Methods for Microbiome Analysis

# Statistical Methods for Microbiome Analysis: A brief review

M. Bhattacharjee, School of Mathematics and Statistics, University of Hyderabad, India

**Corresponding author:**
M. Bhattacharjee, School of Mathematics and Statistics, University of Hyderabad, Hyderabad, India, Email: mbsm@uohyd.ac.in

**Abstract:**
Recent attacks of various viruses with having deep and extensive impact at a global scale has warranted that microbiome be studied extensively and in a robust analytic framework. Microbiome typically refers to the collective genomes of such organisms, although it could also refer to the collection of the organisms by themselves. Here we provide an overview of statistical techniques that are useful in analysing such data.

**Introduction:**
Microscopic living organisms like viruses, bacteria and fungi are referred to as microbes. Recent years have seen multiple attacks of various microbes on human and other living organisms, a prime example being the pandemic caused by COVID19. Microbiome typically refers to the collective genomes of such organisms, although it could also refer to the collection of the organisms by themselves.

This has drawn attention to the necessity of applications of analytic methods that would enable scientists and decisions makers to draw robust conclusions from genomic data on microbes and microbiomes. However, it should be noted that requirement of rigorous analytic techniques for such data have been recognised for a long time [1].

The two main technologies used to gather such genomic data would be 1) 16S ribosomal RNA sequencing (16S rRNA) and 2) meta-genomic shotgun sequencing (MSS). With both these processes, are associated many technical issues, in terms of sequencing technology, the complex nature of sequencing count data, and choice of data normalization techniques. A discussion on these is beyond the scope of this article [5].

The final output of most microbiome analysis pipelines is an OTU (Operational Taxonomic Units) table. These OTUs are taken as the pragmatic proxies for potential microbial species represented in a sample. Irrespective of which technology and pipeline is used, one has to assure high quality reads along with stringent screening and filtering criteria. This could potentially reduce the number of spurious OTUs.

Then the core step in microbiome analysis is the taxonomic classification of the representative sequences and clustering of OTUs. This phase would also require a phylogenetic annotation of the samples or taxa using a pre-built tree. Using this, one would carry out sequence classification, OTU clustering, and taxonomy assignment.

The next phase would involve, (a) identifying OTU abundance (that is, share of any specific microbe), (b) alpha-diversity (that is, a measure of microbiome diversity applicable to a single sample), and (c) beta diversity (that is, a measure of the similarity or dissimilarity of two communities).

We present summary/outline of methods used in downstream analysis that follow the above path. These methods become critical in answering questions related to phenotypes (such as COVID and non-COVID), covariates (such as age, clinical history, etc.) and microbes (as ascertained by the techniques mentioned above).

**Discussion**
The typical questions or hypothesis that are addressed in microbiome analyses are 1) association of a microbiome with the host, 2) association of microbiome with the environmental covariates and 3) association between the environment and the host.

To compare the diversities between two groups, we use two tests: the parametric test used is the **Two-Sample Welch's t-Test** [2]; its non-parametric counterpart is the **Wilcoxon Rank Sum Test** [2] (or equivalently the **Mann Whitney U-statistic**). The Welch's t-test, which is also known as the unequal variances t-test, is a two-sample location test that is used to test the hypothesis that, the two groups have equal means, with the underlying assumption of Gaussianity for both the groups. The Wilcoxon rank-sum test (which has an exact function relation with the Mann Whitney U statistics-based test) is a nonparametric alternative to the two-sample t-test. It is based solely on the ordering in magnitude of the observations from the two groups in the combined sample.

One would not always be interested in the large spectrum of microscopic organisms, but instead, a specific taxon (or species) may be of importance in a given situation. In attempting to identify a specific species, one has to keep in mind that, from the abundance of taxa in the samples, it is inappropriate to draw conclusion regarding the total abundance in the ecosystem. Nevertheless, we can use the relative abundance in the sample to infer about the relative abundance of a taxon in the ecosystem. This relative abundance can also be assessed using the **Wilcoxon rank-sum test**. Another option would be to use a **Chi-square test,** which is applied to sets of categorical data to test whether an observed frequency distribution differs from a claimed or theoretical distribution. Interestingly, a **Chi-square test** [1] may also be used to assess independence--that is to assess for sets of categorical data, how likely it is that the observed difference between the sets arose merely by chance. A more general test would be **Fisher's exact test** [2] which is useful for two-way classified data and for assessing association between the two classifications.

Generalization of the above ideas would lead us to comparisons for more than two groups. The parametric approach would be to use an **ANOVA** [2] method; and its non-parametric counterpart would be a **Kruskal-Wallis test** [2]. A one-way-analysis of variance or ANOVA facilitates formal assessment of whether or not the means of two or more independent groups are identical. ANOVA relies crucially on the assumption of Gaussianity for the error term in the corresponding linear model. In Kruskal Wallis test the null hypothesis is that the mean ranks of the groups are the same (instead of the means of the actual observations).

Two further techniques related to the multi-group comparison scenario are 1) correction for multiple testing and 2) post-hoc analyses. The **Benjamini–Hochberg** [3] method controls the False Discovery Rate (FDR), using a sequential modified Bonferroni correction for multiple hypothesis testing. **Tukey's HSD test** [2], is a single-step multiple comparison procedure to identify means that are significantly different from each other. A non-parametric tool for the same is **Dunn's multiple comparisons test** [2]**,** which compares the difference in the sum of ranks between two columns with the expected average difference (based on the number of groups and their sizes). The **Newman–Keuls or Student–Newman–Keuls** (SNK) method, is a stepwise multiple comparisons procedure used to identify means that are significantly different from each other. Unlike Tukey's range test, the Newman–Keuls method uses different critical values for different pairs of mean comparisons. The **Dwass–Steel–Critchlow–Fligner tests** is another non-parametric test that compares the median/means of all pairs of groups, using the Steel-Dwass-Critchlow-Fligner pairwise ranking nonparametric method, and it also controls the error rate simultaneously.

Most of the hypotheses described above are focussed and/or specific. It is possible to address/assess more general questions like how to identify which microbiomes significantly contributed to the classification. To address this, the **Orthogonal partial least squares discriminant analysis** (OPLS-DA) [4] has been proposed in literature. In OPLS-DA, a regression model is constructed between the multivariate data and a response variable that only contains class information. This model allows one to eliminate the effect of inter-subject variability among the participants, and identifies microbiomes that significantly contributed to the classification. In addition, it is also possible to rank the microbiomes using VIP or **variable importance in the projection** technique. Many authors use well known Linear discriminant analysis with modifications for assessing effect size. LEfSe (**Linear discriminant analysis Effect Size**) [4] determines the features (organisms, clades, operational taxonomic units, genes, or functions) most likely to explain differences between classes, by coupling standard tests of statistical significance with additional tests which encode biological consistency and effect relevance.

A key question for microbiome community is to study significance of the microbial diversity difference between groups that requires beta diversity indices. A nonparametric extension of multivariate ANOVA, called **Permutational multivariate analysis of variance** (PERMANOVA) [3] enables assessment of microbial diversity difference between groups. A distance measure is required in order to use the PERMANOVA method. For community microbiome data, the **Bray-Curtis distance** [3] has been found to be more appropriate. Other distance measures that are available in the literature include, the **Jaccard distance** [3], and the weighted and unweighted **UniFrac distance** [3].

In recent years interest has grown in model-based approaches to community level analysis of the microbiome. Some of these are developed in a **multivariate generalized linear (mixed) model** framework that includes fully specified joint model for all taxon abundances and related to covariates or phenotypes. For microbiome data, such joint modelling of abundance data can account for the mean–variance relationship and overdispersion. When the data used is in count data form, then several **random effects models** and (hierarchical) **Bayesian models** [3] have been proposed in literature of this purpose.

**Dirichlet multinomial mixtures** (DMM) [3] is also used for probabilistic modelling of the microbial metagenomics data. This data can be represented as a frequency matrix giving the number of times each taxon is observed in each sample. The samples have different size, and the matrix is sparse, as communities are diverse and skewed to rare taxa. In DMM each community is represented by a vector of taxa probabilities. These vectors are generated from one of a finite number of Dirichlet mixture components each with different hyperparameters. Observed samples are assumed to be generated through multinomial sampling.

**Generalised linear models** (GLMs) [3], and their extensions, that are suitable to account for zero-inflated nature of the microbiome data include **zero-inflated beta regression** [3] model for proportional data and **negative binomial regression** [3] model for count data. Negative binomial model can be further modified to **hurdle negative binomial model** [3] to account for excessive zeros.

**Conclusion**
There are two main issues with the methods that are typically seen to be used in the various microbiome data analysis publications. First, a lack of understanding of what are parametric/non-parametric models. Second, a lack of appreciation of some critical complexities of microbiome data.

Regarding the first, many articles and some books refer to parametric methods as those based on Gaussian assumption. However Gaussian models form only a small part of parametric models. For example, most commonly used regression models, including ANOVA, multiple regression, logistic regression, generalised linear model, mixed effects models, are all examples of parametric models. Many of these techniques do not require the Gaussian assumption. While parametric models possess many desirable properties, they rely on the appropriateness of the distributional assumptions that are made. If there is no clear probability law which is supported by the data, then perhaps nonparametric methods should be used. It should also be noted that, both parametric and non-parametric methods are vulnerable to a small sample size—the non-parametric methods more so than the parametric methods.

The other aspect of microbiome data is that, they are naturally high dimensional. In addition, a typical OTU table would produce data type known as "compositional data". These are special type of data that require specific analytic techniques, especially when we desire to study the entire spectrum of taxa present in the data. Additionally, OTU data is commonly sparse in nature, with large proportion of zero counts. Specialized models for handling such excessive zero counts are needed. Here too, we have noticed that many authors are unaware that models for over-dispersion can also be used for such data.

In all, there is need for advanced modelling techniques for complex and large data such as those arising from microbiome studies, and there is a lot to be explored in terms of analyses of such data.

**Conflict of Interest**
No conflict of interest.